\documentclass[aps,prd,preprint,tightenlines,preprintnumbers,showpacs,superscriptaddress,nofootinbib,eqsecnum,floatfix
]{revtex4-1}

\usepackage{bm}
\usepackage{graphicx}
\usepackage[pdftex]{color}
\usepackage{amsmath, amssymb}
\usepackage{hyperref}
\usepackage{dcolumn}
\usepackage{array}
\hypersetup{
    colorlinks=true,     
    linkcolor=blue,      
    citecolor=blue,      
    filecolor=blue,      
    urlcolor=blue        
}
\usepackage{rotating}
\usepackage{mathtools}
\usepackage{multirow}
\usepackage{slashed} 

\definecolor{red}{rgb}{0.8,0.0,0.0}
\definecolor{green}{rgb}{0.0,0.6,0.0}
\definecolor{darkblue}{rgb}{0.0,0.1,0.7}
\definecolor{brown}{rgb}{0.6,0.1,0.0}
\definecolor{gray}{rgb}{0.6,0.6,0.6}
\definecolor{darkgreen}{rgb}{0.0, 0.545098, 0.0}
\definecolor{verydarkgreen}{rgb}{0.0, 0.4, 0.0}
\definecolor{veryverydarkgreen}{rgb}{0.0, 0.3, 0.0}
\definecolor{purple}{rgb}{0.5,0.0,0.5}
\definecolor{applegreen}{rgb}{0.55, 0.71, 0.0}
\definecolor{babypink} {rgb}{0.64, 0.44, 0.44}
\definecolor{orange}{rgb}{0.9,0.4,0.0}

\newcommand{\bi}{\begin{itemize}}
\newcommand{\ei}{\end{itemize}}

\newcommand{\be}{\begin{equation}}
\newcommand{\ee}{\end{equation}}

\newcommand{\bea}{\begin{eqnarray}}
\newcommand{\eea}{\end{eqnarray}}

\newcolumntype{P}[1]{>{\centering\arraybackslash}p{#1}} 

\def\Eq#1{Eq.~(\ref{#1})}
\def\Eqs#1{Eqs.~(\ref{#1})}

\newcommand{\mcal}[1]{\mathcal{#1}} 
\newcommand{\abs}[1]{\left| #1 \right|} 

\newcommand{\spose}[1]{\hbox to 0pt{#1\hss}}
\newcommand{\inapprox}{\mathrel{\spose{\lower 3pt\hbox{$\mathchar"218$}} \raise 2.0pt\hbox{$\mathchar"232$}}}

\newcommand{\inv}{^{-1}} 

\DeclareMathOperator{\arcosh}{arcosh}

\newcommand{\chisq}{\ensuremath{\chi^2}} 

\newcommand{\tmin}{\ensuremath{t_\text{min}}}
\newcommand{\tmax}{\ensuremath{t_\text{max}}}

\newcommand{\ket}[1] {|#1\rangle} 
\newcommand{\matrixel}[3]{\left\langle #1 \right| #2 \left| #3 \right\rangle} 

\newcommand{\Dslash}{\ensuremath{D\kern-0.6em/\kern0.15em}} 
\newcommand{\Tr}{\ensuremath{\mathop{\text{Tr}}}} 

\newcommand{\MSbar}{\ensuremath{\overline{\rm MS}}} 

\newcommand{\GeV}{~\ensuremath{\text{GeV}}}   
\newcommand{\MeV}{~\ensuremath{\text{MeV}}}  
\newcommand{\fm}{~\ensuremath{\text{fm}}} 

\newcommand{\EW}{\ensuremath{\text{EW}}} 
\newcommand{\HC}{\ensuremath{\text{HC}}} 
\newcommand{\EHC}{\ensuremath{\text{EHC}}} 

\def\SU{{\rm SU}}
\def\SO{{\rm SO}}

\def\U1{{\rm U}(1)}

\newcommand{\4}{\ensuremath{\text{\textbf{4}}}} 
\newcommand{\6}{\ensuremath{\text{\textbf{6}}}} 

\def\ha{\hat{a}}

\def\hm{\hat{m}}

\begin{document}


\title{Partial compositeness and baryon matrix elements on the lattice}



\author{Venkitesh Ayyar}
\affiliation{Department of Physics, University of Colorado, Boulder, Colorado 80309, USA}
\affiliation{NERSC, Lawrence Berkeley National Lab  Berkeley, CA 94720 USA}

\author{Thomas DeGrand}
\affiliation{Department of Physics, University of Colorado, Boulder, Colorado 80309, USA}

\author{Daniel C.~Hackett}
\affiliation{Department of Physics, University of Colorado, Boulder, Colorado 80309, USA}

\author{William~I.~Jay}\email{wjay@fnal.gov}
\affiliation{Theoretical Physics Department, Fermi National Accelerator Laboratory, Batavia, Illinois, 60510, USA}
\affiliation{Department of Physics, University of Colorado, Boulder, Colorado 80309, USA}

\author{Ethan T.~Neil}\email{ethan.neil@colorado.edu}
\affiliation{Department of Physics, University of Colorado, Boulder, Colorado 80309, USA}
\affiliation{RIKEN-BNL Research Center, Brookhaven National Laboratory, \\ Upton, New York 11973, USA}

\author{Yigal~Shamir}
\affiliation{Raymond and Beverly Sackler School of Physics and Astronomy,
Tel~Aviv University, 69978 Tel~Aviv, Israel}

\author{Benjamin Svetitsky}
\affiliation{Raymond and Beverly Sackler School of Physics and Astronomy,
Tel~Aviv University, 69978 Tel~Aviv, Israel}

\date{\today}

\begin{abstract}
Partial compositeness is a mechanism for the generation of fermion masses which replaces a direct Higgs coupling to the fermions by a  linear mixing with heavy composite partners.
We present the first calculation of the relevant matrix element in a lattice model which is very close to a candidate theory containing a composite Higgs boson and a partially composite top quark. Specifically, our model is an \SU(4) gauge theory coupled to dynamical fermions in the fundamental and two-index antisymmetric (sextet) representations.
The matrix element we obtain is small and hence our result disfavors the scenario of obtaining a realistic top mass in this model.
\end{abstract}

\preprint{FERMILAB-PUB-18-676-T}
\maketitle
\tableofcontents

\begin{flushleft}


\end{flushleft}


\section{Introduction \label{sec:intro}}

Partial compositeness was introduced by Kaplan~\cite{Kaplan:1991dc} as a method for generating fermion masses via linear coupling to heavy fermionic states in a new composite sector.
In this paper we use lattice gauge theory to study this mechanism in an \SU(4) gauge theory with dynamical fermions in two representations, the fundamental \4 and the two-index antisymmetric \6.
This theory is a slight modification of an asymptotically free model due to Ferretti \cite{Ferretti:2013kya,Ferretti:2014qta,Ferretti:2016upr}, which contains a composite Higgs boson and a partially composite top quark.

Our group has considered other aspects of this model in previous work, including its meson and baryon spectrum and its thermodynamic properties \cite{Ayyar:2017qdf,Ayyar:2018ppa,Ayyar:2018zuk}.
The present work is our first to consider the mixing aspects of partial compositeness.
The particular focus of this paper is a baryon matrix element which, within certain approximations, appears in the formula for the top quark's effective Yukawa coupling and mass.
This work is the first lattice study of partial compositeness in a realistic model.

Our main conclusion is that it is unlikely that partial compositeness gives the top quark a realistic mass in this model.
This stems from the smallness of the calculated matrix element.
The result depends on two key approximations.
First, we change the numbers of flavors of the two species of fermions compared to Ferretti's model.
Second, we relate the top Yukawa coupling to the baryon matrix element by saturating the relevant low-energy constant with the lightest baryon intermediate state.
We believe that improving on these approximations will not supply the orders of magnitude needed to make the model viable.

The outline of the paper is the following.
Section~\ref{sec:pcf} reviews partially composite fermions and the physical context for the nonperturbative calculation.
Section~\ref{sec:lattice} describes the details of the lattice simulation.
Section~\ref{sec:discussion} summarizes our results.
Technical details appear in the appendices.

For lattice work on a different composite Higgs model, see
Refs.~\cite{Bennett:2017kga,Lee:2018ztv}.
For additional related phenomenological work,
see Refs.~\cite{Bizot:2016zyu,Belyaev:2016ftv,DelDebbio:2017ini,Agugliaro:2018vsu}.

\section{Partially Composite Fermions \label{sec:pcf}}

\subsection{The physical picture}
Partial compositeness generates fermion masses through linear coupling to heavy partner states.
In principle, any number of the fermions in the Standard Model could acquire their mass through this mechanism.
Guided, however, by the observation that the top quark is the only fermion in the Standard Model with its mass at the weak scale, we follow a common  practice \cite{Bellazzini:2014yua,Panico:2015jxa} and single it out.
Partial compositeness involves three energy scales:
\begin{enumerate}
\item the low-energy scale $\Lambda_\EW$ of the electroweak (EW) sector of the Standard Model, characterized by the masses of the Higgs, $W$, and $Z$ bosons and of the top quark;
\item an intermediate scale $\Lambda_\HC$, perhaps a few TeV, associated with a new confining ``hypercolor'' (HC) dynamics; and
\item a high-energy scale $\Lambda_\EHC$ of an ``extended hypercolor" (EHC) dynamics, associated with operators needed to generate Standard Model fermion masses.
\end{enumerate}
We shall assume that these energy scales are well separated:
$\Lambda_\EW \ll \Lambda_\HC \ll \Lambda_\EHC$.
The setup is reminiscent of the Standard Model itself, where quantum electrodynamics, hadronic physics, and electroweak physics enjoy large separations of scale:
$\Lambda_\text{QED} \ll \Lambda_\text{QCD} \ll \Lambda_\text{EW}$.
At each scale, this separation allows for an effective field theory description of the dynamics resulting from all the higher scales.  Our SU(4) gauge theory is the hypercolor theory at the scale $\Lambda_\HC$.

This scenario contains four principal ingredients.
First is the fundamental top quark field, which starts out massless.
At low energies, the Standard Model adequately describes its interactions, and the familiar formula, $m_t = y_t v$, furnishes its mass.
If the Standard Model stands alone, both the Higgs vacuum expectation value $v$ and the top Yukawa coupling $y_t$ are parameters which must be determined from experiment.

Second enters a composite Higgs boson.
The present scenario imagines the Higgs to be a Goldstone boson of the hypercolor theory, which confines and spontaneously breaks its chiral symmetry.
Its vacuum expectation value and mass are calculable in terms of low-energy constants in an effective theory after perturbative coupling to the Standard Model.
The fact that the Higgs is now composite provides a solution to the naturalness problem.

Third, the confining hypercolor theory produces bound states with the same quantum numbers as the top quark.
The masses of these new baryon states are fully calculable within the hypercolor theory, just as the proton mass is calculable in QCD.
The overall scale $\Lambda_\HC$ must be determined from experiment.

Fourth and finally, we have the extended hypercolor sector.
For the present discussion, its precise dynamics and particle content remain unspecified.
Partial compositeness only requires that, at the intermediate hypercolor scale, it induces effective four-fermion interactions that couple the top quark to its baryonic partners.

With all four pieces in place, the heavy partners of the top quark may be integrated out.
At the low-energy scale $\Lambda_\EW$ this generates an effective interaction between the top quark and the Higgs boson, which reduces to the Yukawa coupling of the Standard Model in the appropriate limit.
The structure of the effective interactions is constrained by symmetry considerations, while the low-energy constants depend on the masses and interactions of the top-partner states, and, in particular, on the four-fermion interactions.

Schematically, matching between the physical descriptions at the hypercolor scale and at low energy trades the top Yukawa coupling $y_t$ for two analogues of Fermi's constant $G_F$.
We call these new couplings $G_L$ and $G_R$, since they multiply the linear coupling of the top quark to hypercolor baryon operators of definite chirality.
As with Fermi's constant in the Standard Model, they would be calculable once the UV-complete theory has been specified.
It should be noted, however, that the problem of writing down a realistic extended hypercolor theory remains unsolved.
In brief, a successful solution would need to overcome many of the same challenges faced by grand unified theories, such as evading anomalies while uniting quarks together with fermions of the hypercolor theory into bigger representations of the EHC gauge force.

Appendix~\ref{app:matching} describes the matching of the effective low-energy theory at the electroweak scale to the hypercolor theory in Ferretti's model~\cite{Golterman:2015zwa,Golterman:2017vdj}.
Resulting from the calculation are the following expressions for the Yukawa coupling and mass of the top quark,
\begin{align} \label{eq:top_mass}
y_t & \approx\ G_L G_R \frac{Z_L Z_R}{M_B F_6}\ , \\
m_t & \approx\  y_t v \ . \label{eq:top_myv}
\end{align}
Here $M_B$ is the mass of the top partner and $F_6$ is the decay constant associated with the composite Higgs boson.
The factors $Z_L$ and $Z_R$ are defined in terms of matrix elements that describe the overlap of the four-fermion operators with a top partner state.
They arise in \Eq{eq:top_mass} after making the approximation that the relevant low-energy constant, defined in terms of a top-partner two-point function, is saturated by the lowest state.
In the present framework, the familiar formula for the top mass, $m_t=y_t v$, is not an identity but is recovered in the approximation where $\sin(v/F_6)  \approx  v/F_6$, which is supported by experimental constraints~\cite{Bellazzini:2014yua,Panico:2015jxa,Aad:2015pla}.

In general, the four-fermi Lagrangian contains several independent couplings for the left-handed top as well as for the right-handed top.
As a result, the top Yukawa coupling is given by a sum of terms of the form of \Eq{eq:top_mass}.

Mass generation through partial compositeness is somewhat similar to the well-known see-saw mechanism, where a massive state (here the hypercolor baryon) couples linearly to a massless state (here the top quark), thereby generating a mass for the latter.
There is one important difference, though.
While the baryons of the hypercolor sector are indeed massive from the outset, the top quark can receive a mass only after the Higgs field develops an expectation value.
Instead of directly generating a mass for the top, the linear coupling between the top and the hypercolor sector generates a Yukawa coupling for the top quark.

\subsection{The scope of the lattice calculation}

Equation~(\ref{eq:top_mass}) gives the mass of the top quark in terms of physical observables in the hypercolor and extended hypercolor sectors.
The couplings $G_{L,R}$ are external to the lattice calculation and are not calculable without a specific UV completion.
All the other factors are calculable on the lattice in terms of dimensionless ratios, given a concrete hypercolor theory like Ferretti's model.
Our group has previously calculated the mass of the top partner $M_B$ and the pseudoscalar decay constant $F_6$ \cite{Ayyar:2017qdf,Ayyar:2018zuk}.

In this work we compute the normalization factors $Z_{L,R}$.
Each four-fermion interaction couples a third-generation quark field to a hypercolor-singlet three-fermion operator, which serves as an interpolating field for the (left-handed or right-handed) top partner.
$Z_{L,R}$ are defined from the matrix elements
of the interpolating fields between the vacuum and a single top partner state.
A lattice-regulated matrix element is converted into a continuum-regulated ($\MSbar$) matrix element at a reference scale, which, in turn, is defined in terms of $\Lambda_\HC$.
We will take $\Lambda_\HC \equiv F_6$ to define the characteristic scale of the hypercolor theory.

\subsection{Symmetries and the top partner \label{sec:symmetries}}

We now discuss the symmetries of Ferretti's model and of ours.
Let $N_4$ and $N_6$ denote the number of flavors of Dirac fermions in the fundamental and sextet representations, respectively.
The fundamental representation is complex, while the sextet is real.
In the present study $N_4 = N_6 = 2$, to be compared with $N_4=3$ and $N_6=5/2$ in Ferretti's model
(that is, Ferretti's model has five Majorana fermions in the \6 representation).
The global symmetry group in the massless limit is $\SU(2 N_6)$ in the sextet sector and $\SU(N_4)_L \times \SU(N_4)_R \times \U1_B$ in the fundamental sector, where $\U1_B$ is the fermion number of the fundamental fermions.
In addition, the model contains a non-anomalous $\U1_A$ axial symmetry.
After spontaneous breaking of chiral symmetry, the unbroken symmetry group is $\SO(2N_6) \times \SU(N_4)_V \times \U1_B$.
Ref.~\cite{Ayyar:2017qdf} discusses some  phenomenological consequences related to the fact that the
\6 representation of \SU(4) is real, while Ref.~\cite{DeGrand:2015lna} contains additional group theoretical details.

We have changed the flavor numbers from Ferretti's model to simplify the lattice calculation.
We do not expect the matrix elements we compute to change significantly from this simplification.
The situation is similar to that of QCD, where most matrix elements show only weak dependence on the number of flavors of fermions active in the simulations.
For a related discussion, see Ref.~\cite{Aoki:2016frl}.

In Ferretti's model, the Standard Model's gauge group lies in the unbroken global subgroup $\SO(5)\times \SU(3)_V \times \U1_B$.
The custodial symmetry group of the Standard Model, $\SU(2)_L \times \SU(2)_R \simeq \SO(4)$, is embedded in the unbroken \SO(5).
Because the electroweak symmetry is embedded in the global symmetry of the sextet fermions, these fermions carry electroweak charges.
The fundamental fermions' unbroken $\SU(3)_V$ subgroup is identified with the gauge group of QCD.
Thus, the fundamental fermions carry color charge in the Standard Model.
Because the top quark carries both electroweak and color charges, the top partner must be a fermionic resonance containing both fundamental and sextet fermions.
Specifically, the top partner is made of one sextet and two fundamental fermions, forming a singlet of the SU(4) gauge group.

Our previous work on the baryon spectrum of the $N_4 = N_6 = 2$ theory considered just such mixed-representation objects, referring to them as \emph{chimera baryons} due to their hybrid nature~\cite{Ayyar:2018zuk}.
The chimera states may be classified according to their total spin $J$ and the ``isospin" $I$ of the fundamental fermions.
Moreover, the spectrum of the chimera states invites understanding through analogy with the hyperons of QCD, with the sextet playing the role of a light strange quark.
The top partner in Ferretti's model corresponds to the $(J,I) = (1/2,0)$ chimera baryon.
This state is the analogue of the $\Lambda$ hyperon and gets its spin from the sextet fermion.
Ref.~\cite{Ayyar:2018zuk} offers more group theoretical details relating to this identification.

\section{The Lattice Computation\label{sec:lattice}}

\subsection{Simulation Details}

This study uses ensembles with simultaneous dynamical fermions in both the fundamental \4~and the two-index antisymmetric \6~representation of SU(4), with two Dirac flavors of each.
We use a Wilson-clover action, with normalized hypercubic (nHYP) smeared gauge links~\cite{Hasenfratz:2001hp,Hasenfratz:2007rf}.
The clover coefficient is set equal to unity for both fermion species~\cite{Bernard:1999kc}.
For the gauge field, we use the nHYP dislocation suppressing (NDS) action, a smeared action designed to reduce gauge-field roughness that would create large fermion forces in the molecular dynamics evolution~\cite{DeGrand:2014rwa}.
This study reuses lattices and propagators generated in previous studies.
We therefore refer the reader to these papers for other technical details~\cite{Ayyar:2017qdf,Ayyar:2018zuk}.

Table~\ref{table:ensemble_summary} summarizes some important properties of the nine ensembles used in this study.
Appendix~\ref{app:data_tables} contains more information:
Table~\ref{table:ensembles} gives the parameters of the individual ensembles, while
Table~\ref{table:fermion_masses} provides the measured values of the fermion masses $\hm_4$ and $\hm_6$ and of the Wilson flow scale $t_0/a^2$.
As in our previous studies of this model, hatted variables denote dimensionless quantities constructed by multiplying by appropriate powers of $t_0/a^2$. For instance, $\hm_{4} \equiv (m_{4} a) \times (\sqrt{t_0/a^2})$.

\begin{table}[t]
\centering
\setlength{\tabcolsep}{12pt} 
\begin{tabular}{ l l l }
\toprule
						& min   & max \\
\hline
$t_0/a^2$			& 1.06  & 1.85 \\
$M_{P4}/M_{V4}$	& 0.55  & 0.79 \\
$M_{P6}/M_{V6}$	& 0.47  & 0.73 \\
$M_{P4}L$		& 4.23  & 8.16 \\
$M_{P6}L$		& 4.03  & 8.91 \\
\toprule
\end{tabular}
\caption{
	Summary of basic physical properties of the ensembles used in this study.
	$M_{Pr}$ and $M_{Vr}$ denote the mass of the pseudoscalar and vector mesons, respectively, in the two representations ($r=4, 6$).
	$L$ denotes the spatial extent of the lattice.
	\label{table:ensemble_summary}
	}
\end{table}

\subsection{Correlation Functions} \label{ssec:correlation_functions}

Our goal is to calculate the overlap factors $Z_{L,R}$, which we define according to
\begin{align} \label{eq:zb_definition}
\matrixel{0}{\mcal{O}_{L,R}^\alpha(\bold{0},0)}{\Lambda,\bold{0},\sigma}
	&= Z_{L,R}\, u(\bold{0},\sigma)^\alpha,
\end{align}
where $\ket{\Lambda,\bold{p},\sigma}$ is a top-partner chimera state of definite momentum and spin, and $u(\bold{p},\sigma)^\alpha$ is an on-shell Dirac spinor.
The operators $\mcal{O}_{L,R}(\bold{x},t)$ are listed below.
To extract the lattice regulated version of this amplitude, we conduct joint correlated fits to the
 following time-slice correlation functions:
\begin{align}
C^{PS}_{\pm}(t)
	&= \sum_{\bold{x}} \Tr \left[P_\pm \matrixel{0}{  \mcal{O}_{L,R}(\bold{x},t) \bar{\Lambda}(\bold{0},0) }{0}\right]
	\sim Z_{L,R} Z_\Lambda e^{-M_B |t|}, \label{eq:c_ps}\\
C^{SS}_{\pm}(t)
	&= \sum_{\bold{x}} \Tr \left[P_\pm \matrixel{0}{  \Lambda(\bold{x},t) \bar{\Lambda}(\bold{0},0) }{0}\right]
	\sim Z_\Lambda^2 e^{-M_B |t|} \label{eq:c_ss},
\end{align}
with $Z_{L,R}$, $Z_\Lambda$, and $M_B$ as free parameters.
The mass $M_B$ was computed already in Ref.~\cite{Ayyar:2018zuk}, and we verified that the new operators used in
this study reproduce the masses on each ensemble.
In these expressions, $P_\pm = \frac{1}{2}(1\pm \gamma_4)$ is a parity projection operator and $\Tr$ denotes a trace over the free spinor indices.
In order to isolate the lowest-lying baryon state, we perform the fit to an exponential on $C_+$ for positive times and $C_-$ for negative times; see Appendix~\ref{app:spectro} for details.
$\mcal{O}_{L,R}$ is a point operator, while $\Lambda$ and $\bar\Lambda$ are smeared.
We employ Gaussian smearing on time slices, fixing to the Coulomb gauge before smearing.

$\Lambda$ is the baryon interpolating field.
In analogy with hyperons in QCD, let $u$ and $d$ denote the two different flavors of fundamental fermions and $s$ denote a sextet fermion.
Then
\begin{align} \label{eq:lambda_operator}
\Lambda = 2s(u C \gamma_5 d) + d (u C \gamma_5 s) - u(d C \gamma_5 s)
\end{align}
where $C$ is the charge-conjugation matrix. We use the following shorthand,
\begin{align}
u (d\Gamma s) \equiv \epsilon_{ABCD}u_{\alpha }^A (d_{\beta }^B \Gamma_{\beta\gamma} s_\gamma^{CD}), \label{eq:index_conventions1}\\
s (u\Gamma d) \equiv \epsilon_{ABCD} s_\alpha^{AB} ( u_{\beta}^C \Gamma_{\beta\gamma} d_{\gamma}^{D}), \label{eq:index_conventions2}
\end{align}
with Greek spinor indices and uppercase Latin hypercolor SU(4) indices.
This operator, familiar from baryon spectroscopy in QCD~\cite{Leinweber:2004it}, has quantum numbers $(J,I)=(1/2,0)$ and is chosen to have strong overlap with the $\Lambda$ baryon.

As written, the global flavor structure of \Eqs{eq:index_conventions1} and~(\ref{eq:index_conventions2}) only makes sense for the theory we are simulating and not for the enlarged global symmetry of Ferretti's model.
For the latter, let $q^a$ denote a fundamental fermion with flavor $\SU(3)$ index $a$, and $Q^i$ denote a sextet fermion with flavor $\SO(5)$ index $i$.
With the same spin and hypercolor structure as above, the counterparts of \Eqs{eq:index_conventions1} and~(\ref{eq:index_conventions2})  with manifest flavor transformation properties are
\begin{align}
	q (q\Gamma Q) \equiv \epsilon_{abc} q^b (q^c \Gamma Q^i),\\
	Q (q\Gamma q) \equiv \epsilon_{abc} Q^i ( q^b \Gamma q^c).
\end{align}

For $\mcal{O}_X$, $X=L,R$, we use the following four operators relevant to partial compositeness in this model \cite{Ferretti:2014qta,Golterman:2015zwa,DeGrand:2015yna,Golterman:2017vdj},
\begin{align}
\mcal{O}_X \in
	\begin{cases}
B_X = \frac{1}{2} \left[ P_X s(u C P_R d) - (u \leftrightarrow d) \right], \\
B_X^\prime
    = \frac{1}{2} \left[ P_X s(u C P_{L} d) - (u \leftrightarrow d) \right]. \\
	\end{cases} \label{eq:4_fermion_operators}
\end{align}
The pair of fundamental fermions is antisymmetric in flavor, spin, and hypercolor.
The primed and unprimed operators differ only in the choice of chiral projector $P_{L,R} = \frac{1}{2}(1 \pm \gamma_5)$ inside the diquark.
In this study we do not consider two additional operators discussed in Ref.~\cite{Golterman:2017vdj}. 

Some of the overlap factors are related by symmetry.
Under the usual parity transformation which maps fermions according to $\psi(\mathbf{x},t) \mapsto \gamma_4 \psi(-\mathbf{x},t)$, the operators in \Eq{eq:lambda_operator} and \Eq{eq:4_fermion_operators} transform as
\begin{align}
\Lambda(\mathbf{x},t)		&\mapsto \gamma_4 \Lambda(-\mathbf{x},t) \\
B_{R,L}(\mathbf{x},t)			&\mapsto \gamma_4 B^\prime_{L,R}(-\mathbf{x},t) \\
B^\prime_{R,L}(\mathbf{x},t)	&\mapsto \gamma_4 B_{L,R}(-\mathbf{x},t),
\end{align}
which follow from standard properties of the charge conjugation matrix, $C \gamma_\mu C = \gamma_\mu^T$ and $C^2 = - 1$.
The transformation properties of the operators imply the following transformation for the correlation function:
\begin{align} \label{eq:parity_relation_correlators}
\sum_{\bold{x}} \Tr \left[P_\pm \matrixel{0}{ B_{R,L}(\bold{x},t)  \bar{\Lambda}(\bold{0},0) }{0}\right]
\mapsto
- \sum_{\bold{x}} \Tr \left[P_\pm \matrixel{0}{ B^\prime_{L,R}(\bold{x},t)  \bar{\Lambda}(\bold{0},0) }{0}\right].
\end{align}
Therefore, the overlap factors are related according to $Z_{R,L} = -Z^\prime_{L,R}$.
To improve statistics, our analysis combines correlation functions related by symmetry.
Appendix~\ref{app:spectro} contains more technical details related to this point.
Below we report results for the two independent overlap factors $Z_L$ and $Z_R$ only.

For use in phenomenology, the lattice regulated operator must be converted to a continuum renormalization scheme.
Technical details related to this conversion are covered in Appendix~\ref{app:renormalization}.
In particular, Appendix~\ref{app:renormalization} includes a discussion of operator mixing and our usage of clover fermions.
Using the matching factor from a one-loop calculation following Lepage and Mackenzie~\cite{Lepage:1992xa} gives
\begin{align}
       Z_{L,R}^{\MSbar}\left(\mu=1/a \right)
       = \left[1 + \frac{\alpha_{\MSbar}(q^*)}{4\pi}{\cal Z} \right] Z_{L,R}^\text{lattice}(\mu = 1/a),
\end{align}
where $\alpha$ is the gauge coupling, $q^*$ is a matching scale proportional to $1/a$, and ${\cal Z}$ is a constant.

Throughout the rest of this paper we shall always report the renormalized quantities $Z_{L,R}^{\overline{MS}}(\mu=1/a)$, denoted simply as $Z_{L,R}$.
In line with our usual practice, we define the dimensionless quantities $\hat Z_{L,R}=Z_{L,R}\, t_0^{3/2}$.

\subsection{Numerical results \label{ssec:numerical_results}}

\begin{figure}[t]
\includegraphics[width=1.00\textwidth]{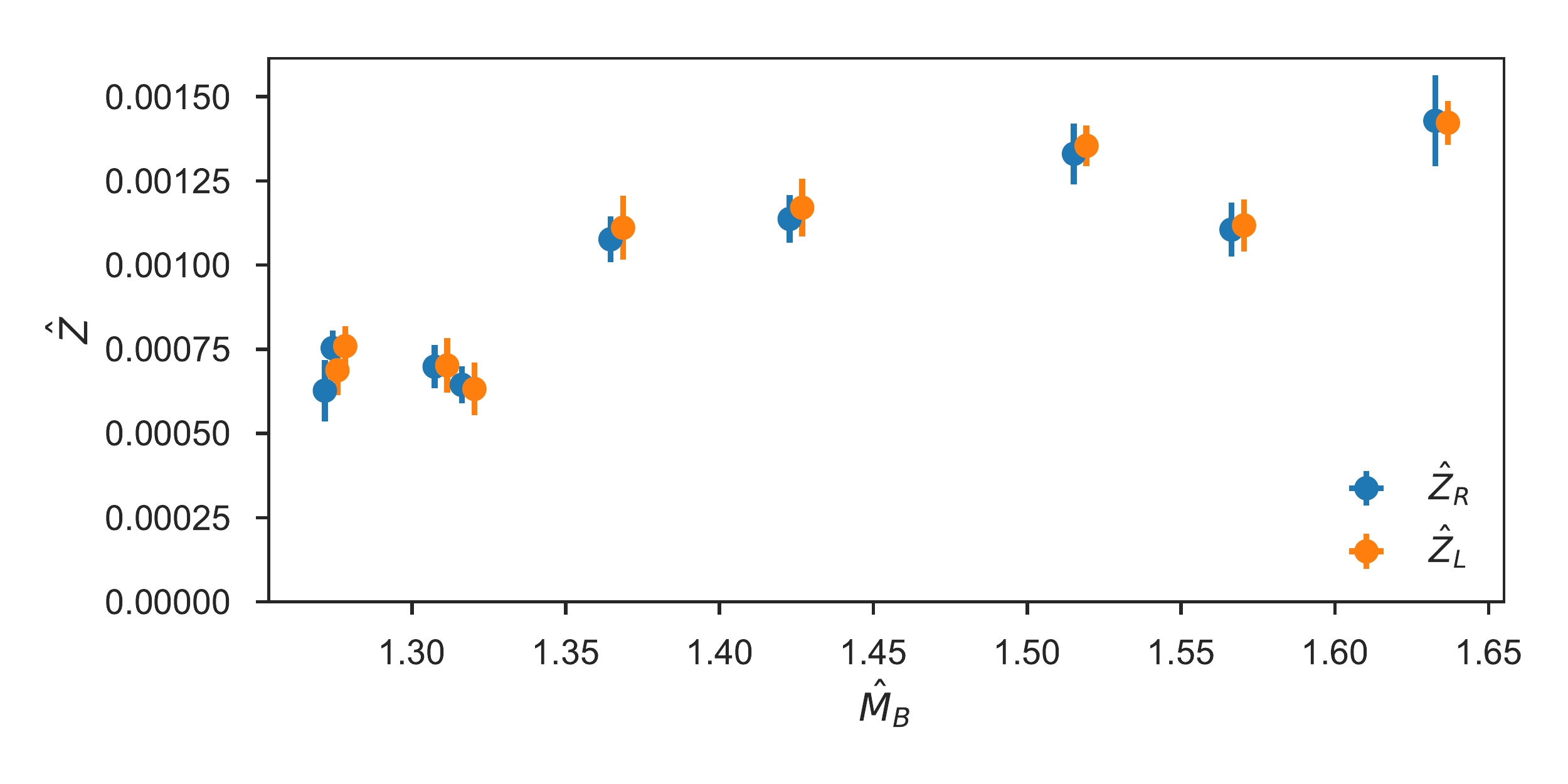}
\caption{
	Lattice data for $\hat{Z}_{L,R}$, the renormalized overlap factors in units of the flow scale $t_0$.
	The data are plotted as a function of the mass $M_B$ of the top partner in the continuum limit, taken from Ref.~\cite{Ayyar:2018zuk}.
	Horizontal positions have been offset slightly to aid readability, and horizontal error bars have been suppressed.
	\label{fig:z_raw_data}
}
\end{figure}

Figure~\ref{fig:z_raw_data} shows our results for $\hat{Z}_L$ and $\hat{Z}_R$, the renormalized overlap factors of the operators $B_L$ and $B_R$ with the state $\ket{\Lambda}$ in units of the Wilson flow scale $t_0$.
Table~\ref{table:raw_overlaps} in Appendix~\ref{app:data_tables}
contains the numerical results themselves.
On each ensemble, these two overlaps are equal within the uncertainties of our computation, in agreement with theoretical expectations~\cite{Golterman:2015zwa}.

The fermion masses, $m_4$ and $m_6$, are free parameters of Ferretti's model.
A sextet Majorana mass term respects the unbroken $\SO(5)$ global symmetry and, therefore, the embedded symmetries of the Standard Model.
Similarly, a fundamental Dirac mass respects the embedded SU(3) color symmetry of QCD.
However, the sextet mass does have a qualitative constraint.
If $m_6$ becomes too large, it will push the global minimum of the Higgs potential back to the origin, thereby obstructing the Higgs mechanism.
Without more detailed quantitative knowledge of the Higgs potential and its low-energy constants, it is hard to specify just how large $m_6$ may safely be.
For this reason, we are most interested in the values of the overlap factors in the continuum limit and when the sextet fermion mass is small.

Although one could imagine fitting these data using heavy baryon chiral perturbation theory, we proceed along more pedestrian lines.
We use a simple four-parameter linear model for the overlap factors $\hat{Z}_{L,R}$:
\begin{align} \label{eq:chiral_ctm_model}
\hat{Z}_{L,R} = p_0 + p_4 \hm_4 + p_6 \hm_6 + p_a \ha.
\end{align}
The raw data motivate this model.
As Fig.~\ref{fig:z_raw_data} suggests, the overlap factors are fairly smooth as a function of the baryon mass, which in turn can be approximated well as a linear function of $\hm_4$ and $\hm_6$~\cite{Ayyar:2018zuk}, albeit with some scatter.
Our previous experience in this \SU(4) model suggests that the residual scatter may be the result of lattice artifacts.
We model this effect through the term linear in the lattice spacing $\ha \equiv a / \sqrt{t_0}$.

\begin{figure}[t]
\includegraphics[width=1.0\textwidth]{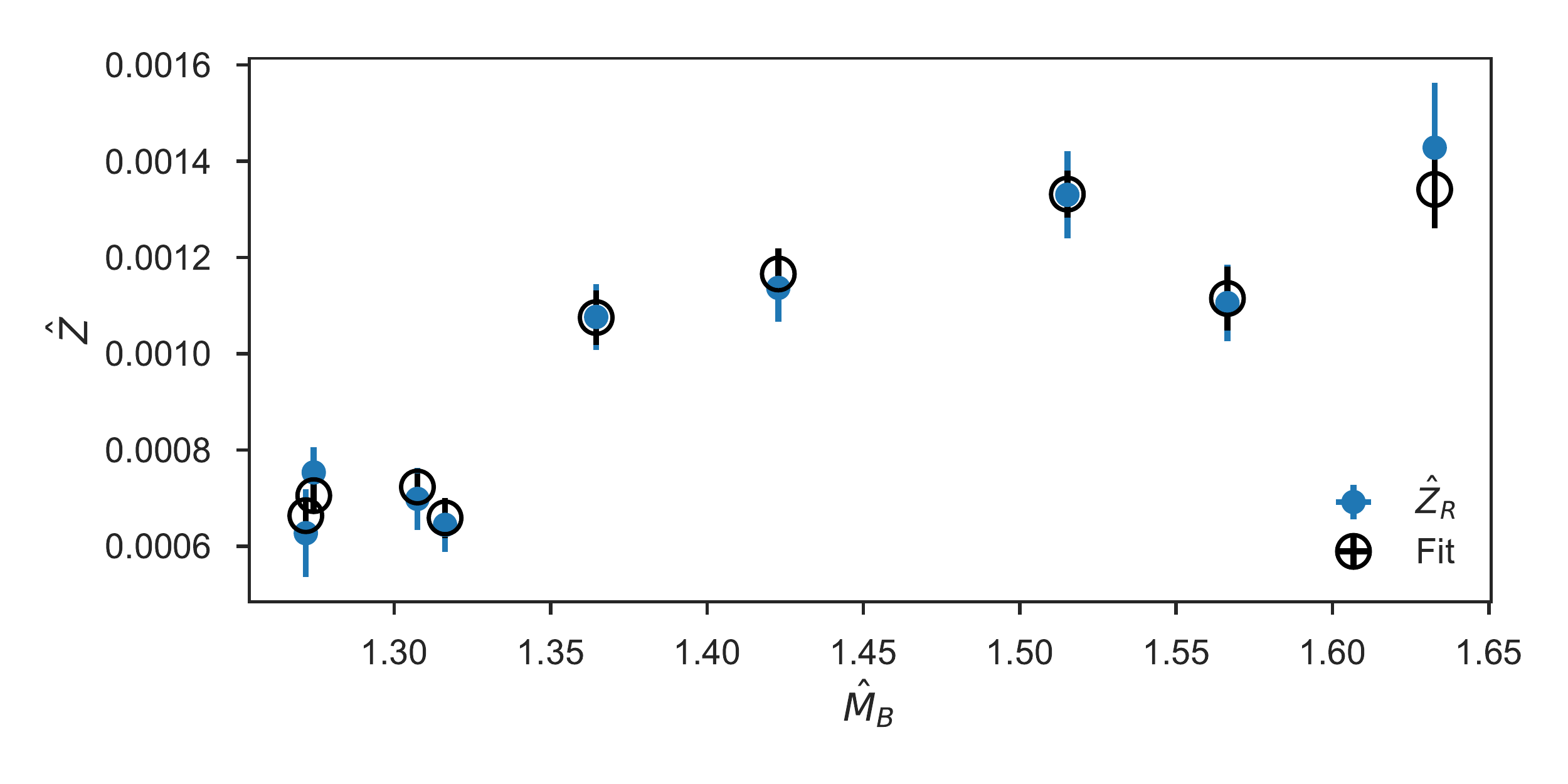}
\caption{
	Fitting the renormalized overlap factor $\hat{Z}_R$ with \Eq{eq:chiral_ctm_model}.
	Solid symbols denote data, while hollow black symbols denote the fit result.
	The corresponding result for $\hat{Z}_L$ is similar.
	The horizontal axis is the same as in Fig.~\ref{fig:z_raw_data}.
	\label{fig:z_raw_data_with_fits}
}
\end{figure}

Figure~\ref{fig:z_raw_data_with_fits} shows the results of fitting $\hat{Z}_R$ to the model in \Eq{eq:chiral_ctm_model}; the fit for $\hat{Z}_L$ is similar.
The fits are successful, with $\chisq=1.85$ and~1.55 for $\hat{Z}_R$ and $\hat{Z}_L$, respectively, each with 5 degrees of freedom.

\begin{figure}[t]
\includegraphics[width=1.0\textwidth]{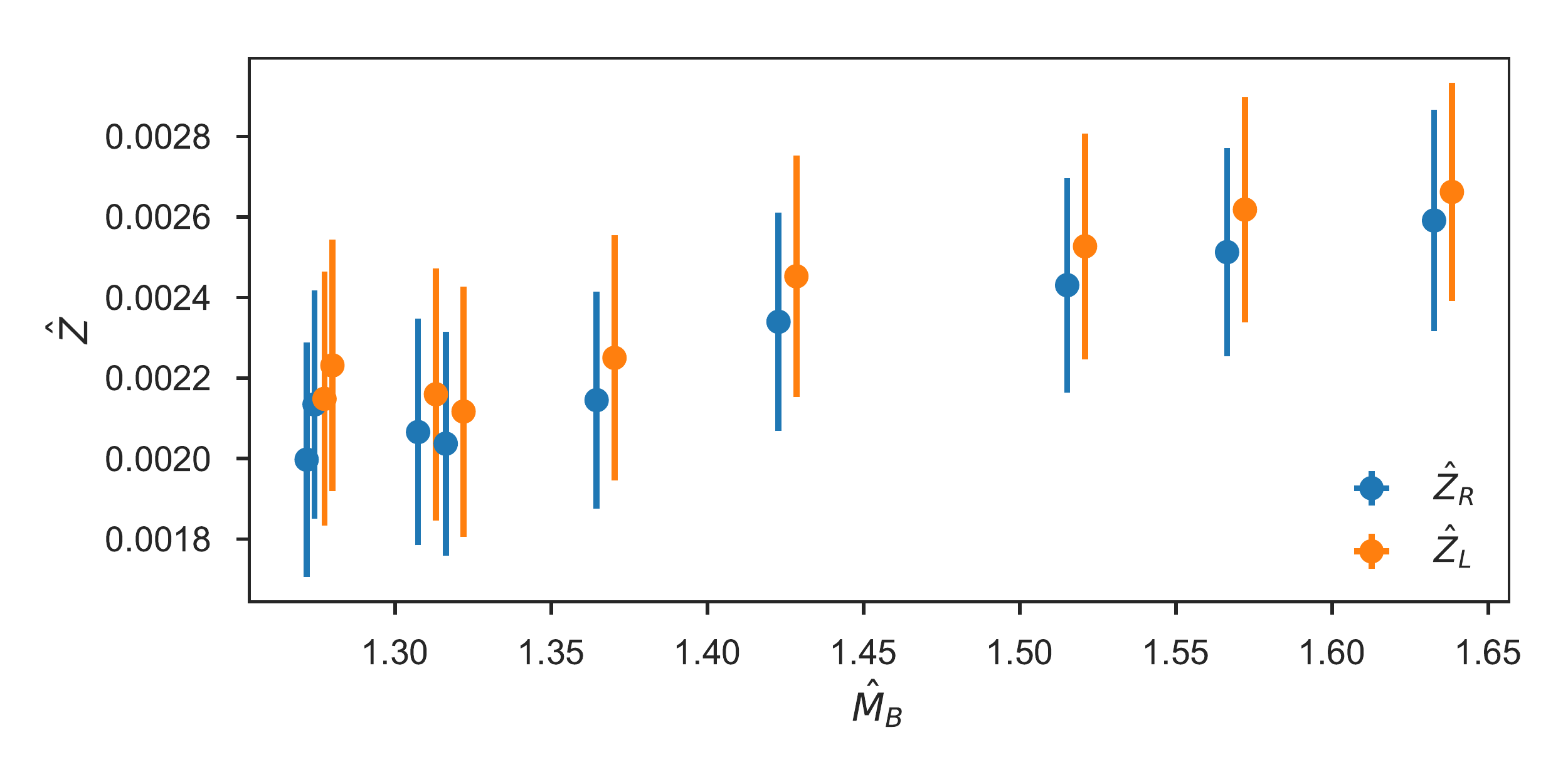}
\caption{
	The overlap factors in the continuum limit.
      The horizontal axis is the same as in Fig.~\ref{fig:z_raw_data}.
	\label{fig:z_t0_units_continuum_limit_only.pdf}
}
\end{figure}

\begin{figure}[t]
\includegraphics[width=1.0\textwidth]{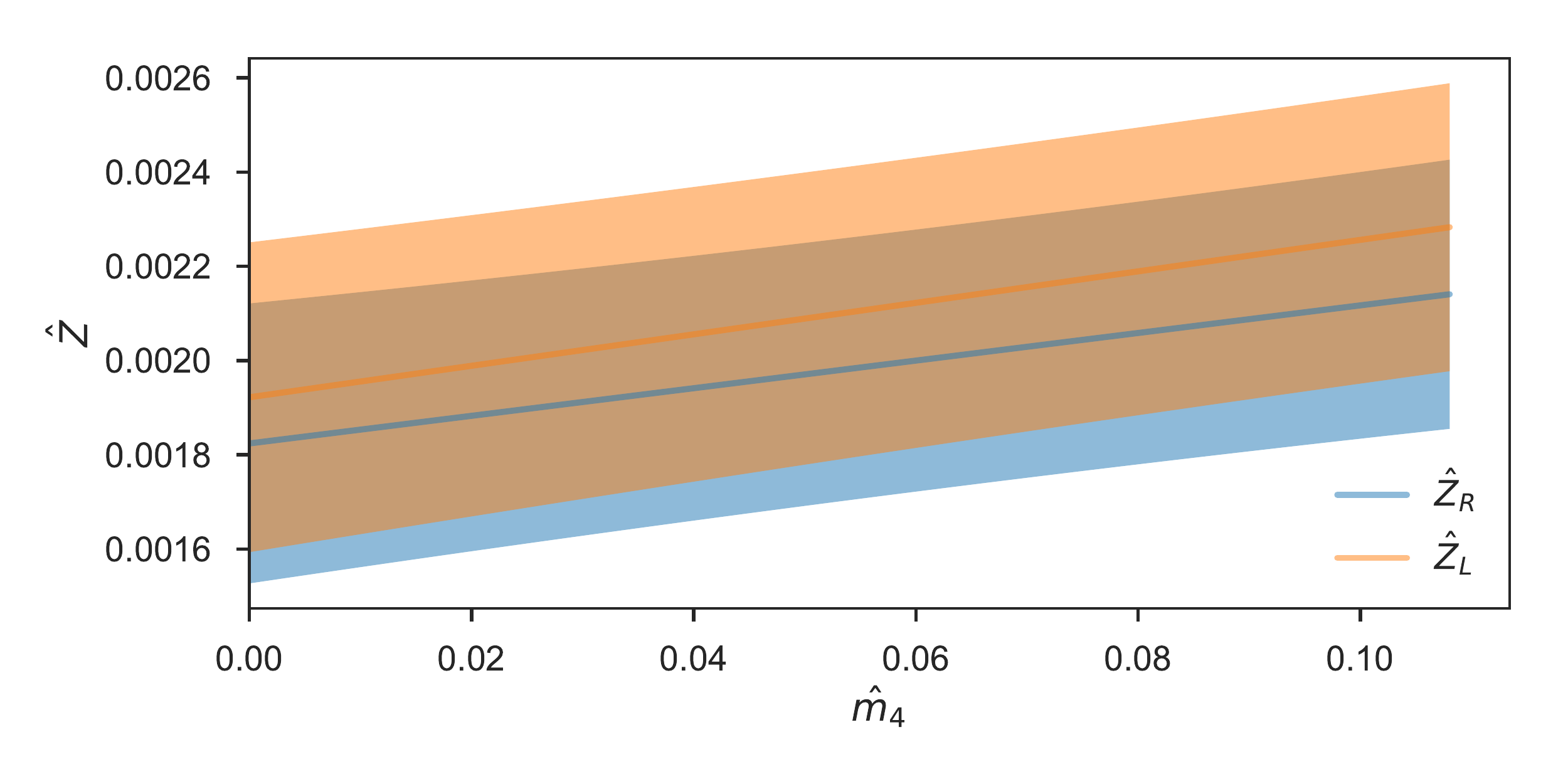}
\caption{
	The overlap factors  in the joint continuum and sextet-chiral ($m_6 \rightarrow 0$) limit, as a function of the fundamental fermion mass $\hat{m}_4$.
	These limits are taken from the fits to \Eq{eq:chiral_ctm_model}.
	\label{fig:z_t0_units_ferretti_limit}
}
\end{figure}

\begin{figure}[t]
\includegraphics[width=1.0\textwidth]{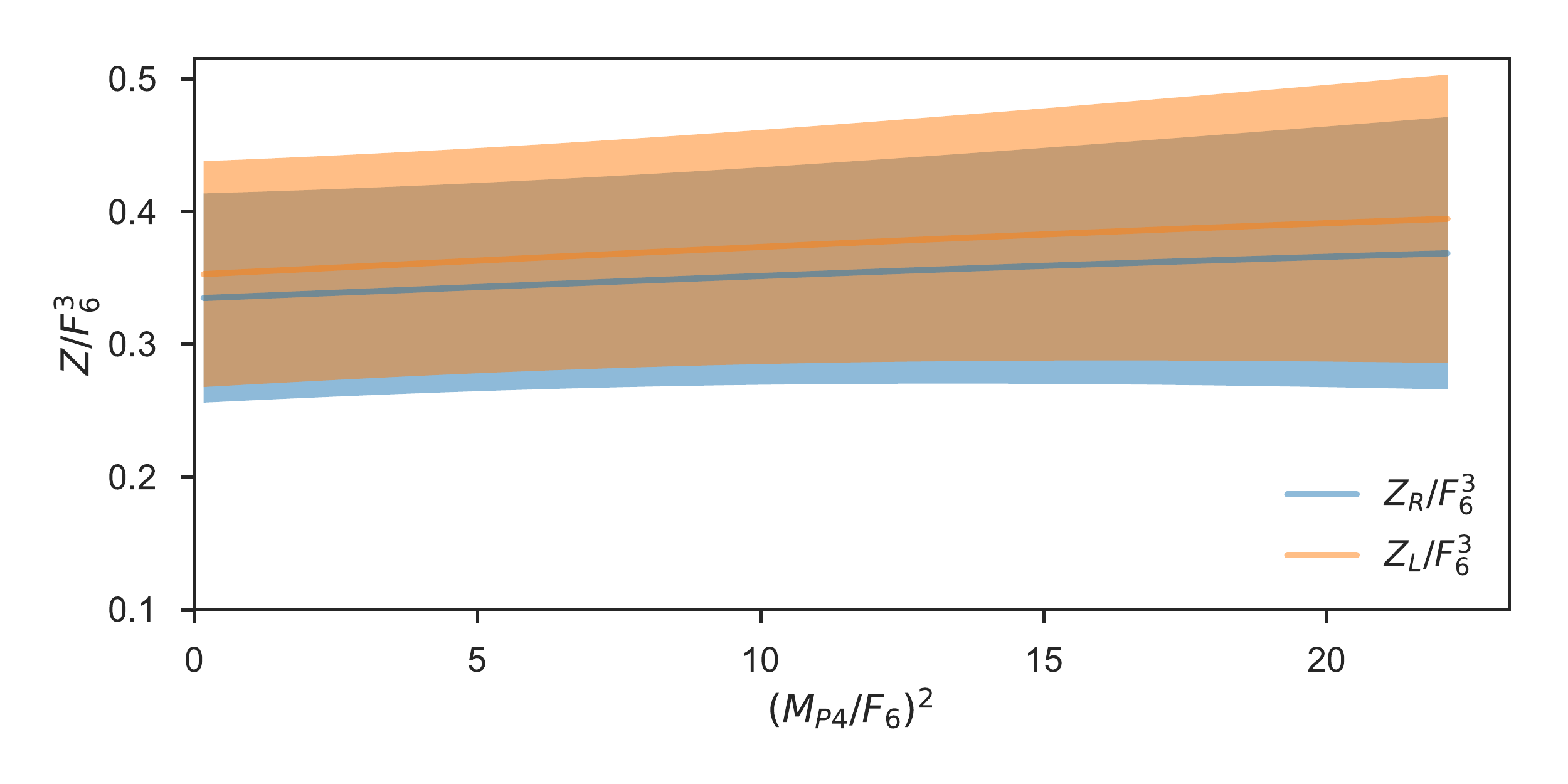}
\caption{
	The overlap factors in the joint continuum and sextet-chiral ($m_6 \rightarrow 0$) limit, plotted against the squared mass of the fundamental pseudoscalar.
	The axis variables are dimensionless ratios constructed with the sextet's pseudoscalar decay constant $F_6$, calculated in Ref.~\cite{Ayyar:2017qdf}.
	\label{fig:z_f_units_ferretti_limit}
}
\end{figure}

We use the fits to construct the overlap factors in various limits.
First, in Fig.~\ref{fig:z_t0_units_continuum_limit_only.pdf} we construct the continuum $(a\rightarrow 0)$ limit by subtracting from the data the lattice artifact identified in the fits.
This lattice artifact is large and negative.
We also use the fits to construct the overlap factors in the simultaneous continuum $(a\rightarrow 0)$ and chiral-sextet $(m_6 \rightarrow 0)$ limit.
Figure~\ref{fig:z_t0_units_ferretti_limit} shows these quantities in units of the Wilson flow scale $t_0$.
Phenomenologists may find the results more interesting as dimensionless ratios with $F_6$, the sextet pseudoscalar decay constant, which we have studied previously~\cite{Ayyar:2017qdf,DeGrand:2016pgq}.
Figure~\ref{fig:z_f_units_ferretti_limit} shows the overlap factors $Z_{L,R} / F_6^3$ as a function of $(M_{P4}/F_6)^2$, a dimensionless proxy for the fundamental fermion mass.
Taken together, Figs.~\ref{fig:z_t0_units_continuum_limit_only.pdf}, \ref{fig:z_t0_units_ferretti_limit}, and~\ref{fig:z_f_units_ferretti_limit} show that the overlap factors are rather flat functions of $m_4$ and $m_6$, the free parameters in Ferretti's model.

Lattice calculations can be affected by (among other factors) the size of the simulation volume.
We have not simulated multiple volumes, so we cannot see this dependence directly.
However, we can compare our volumes to those of QCD simulations if we temporarily set the flow parameter to its QCD value, $\sqrt{t_0} \simeq 0.14 \fm$, and then present our simulation volumes in fm:  $V \simeq (1.6 \fm)^3$--$(2.2 \fm)^3$.
This is similar to the volumes of $(1.8 \fm)^3$ and $(2.7 \fm)^3$ used in a lattice calculation of the analogue quantity in QCD (see below), which saw no noticeable finite-volume effects~\cite{Aoki:2008ku}.
We also note that $M_P L > 4$ for all our data sets (as shown in Table \ref{table:ensemble_summary}).
We therefore have grounds to claim that finite-volume effects are small in our calculation.

\section{Discussion \label{sec:discussion}}

\subsection{Comparison to QCD}

The present results for the overlap factors $Z_{L,R}$ may be compared to QCD studies related to proton decay.
The low-energy effective action of grand unified theories often contains four-fermion operators $\mcal{O}^\slashed{B}$ which violate baryon number~\cite{Buras:1977yy,Weinberg:1979sa,Abbott:1980zj,Wilczek:1979hc}.
Typical proton decay channels appearing in this context include
$p\rightarrow \pi^0 e^+$ and
$p\rightarrow \pi^+ \bar{\nu}_e$.
A common theoretical goal is therefore to compute the matrix elements $\matrixel{\pi}{\mcal{O}^\slashed{B}}{P}$.
Studies of these matrix elements date back more than thirty years and continue to this day; Refs.~\cite{Brodsky:1983st,Falkensteiner:1984bv,Hara:1986hk,Gavela:1988cp,Aoki:2006ib,Aoki:2008ku,Aoki:2017puj} provide a useful but incomplete sampling of the literature.

Direct computation of these matrix elements amounts to computing a three-point correlation function.
Chiral symmetry and soft-pion theorems relate these matrix elements to $\matrixel{0}{\mcal{O}^\slashed{B}}{P}$.
The latter matrix elements are easier to compute on the lattice, requiring only two-point functions.
They are the QCD-analogues of the overlap factors defined in \Eq{eq:zb_definition} above.

How big are the overlap factors in QCD?
In rough physical terms, we expect them to be approximately the square of the proton wave function at the origin.
Dimensional analysis provides an order-of-magnitude estimate,
\begin{equation} \label{eq:dimensional_estimate}
Z \approx \abs{\Psi(0)}^2 \approx \frac{1}{\pi R^3} \simeq 0.005 \GeV^3,
\end{equation}
where $R \simeq 0.8 \fm$ is the radius of the proton.
Models in the early literature typically yielded estimates falling roughly between $0.004 \GeV^3$ and $0.015 \GeV^3$~\cite{Buras:1977yy}.
To our knowledge, the most precise lattice determination of the matrix element in QCD appears in Ref.~\cite{Aoki:2017puj}, where the authors determine that $Z = 0.0144(3)(21) \GeV^3$ at a renormalization scale of $\mu=2 \GeV$ in the \MSbar~NDR scheme.
In terms of dimensionless ratios, their result corresponds to
\begin{equation}
	\left. \begin{aligned}
	Z t_0^{3/2}	& \simeq 0.005 \\
	Z / f_\pi^3 	& \simeq  7,
	\end{aligned} \right\rbrace
	\text{ in QCD,}
\end{equation}
using $\sqrt{t_0} \simeq 0.14 \fm \simeq 0.71 \GeV\inv$ and $f_\pi \simeq 130 \MeV$.

Returning to the present model, the values shown in Fig.~\ref{fig:z_t0_units_ferretti_limit} for $\hat{Z}_{L,R} = Z_{L,R} t_0^{3/2}$ are about 2.5 times smaller than their QCD counterparts, which places them at the lower end of the range estimated in the early literature.
More dramatically, the results for $Z_{L,R}/F_6^3$, shown in Fig.~\ref{fig:z_f_units_ferretti_limit}, are smaller than their QCD counterparts by about a factor of 20.
This has significant phenomenological implications, as we discuss next.

\subsection{Implications for phenomenology} \label{ssec:pheno}

Returning to \Eq{eq:top_mass} and suppressing the $L,R$ subscripts, the top quark Yukawa coupling is, schematically,
\begin{align} \label{eq:ytsim}
y_t \sim \frac{G^2 Z^2}{M_B F_6} \ .
\end{align}
The effective coupling can be expressed as
\begin{align} \label{eq:Gg}
G \sim g_\EHC^2/\Lambda_\EHC^2 \ ,
\end{align}
where the dimensionless coupling $g_\EHC$ characterizes the extended hypercolor dynamics.
If the four-fermion interaction arises from the exchange of weakly coupled heavy gauge bosons, one might expect $g_\EHC^2\sim 0.1$.
Re-arranging terms, we find
\begin{align} \label{eq:fiddle2}
y_t \sim
	\left(\frac{g_\EHC F_6}{\Lambda_\EHC}\right)^4
	\left(\frac{Z}{F_6^3}\right)^2 \frac{F_6}{M_B} \ .
\end{align}
This rearrangement is convenient since we see in Fig.~\ref{fig:z_f_units_ferretti_limit} that $Z/F_6^3 \simeq 0.3$,
and our previous calculation found that $F_6/M_B \simeq 1/6$ \cite{Ayyar:2018zuk}.
The product of the last two factors in \Eq{eq:fiddle2} is about 0.01. As $y_t \simeq 1$, it follows that we need
\begin{align} \label{eq:100}
\left(\frac{g_\EHC F_6}{\Lambda_\EHC}\right)^4 \simeq 100 \ ,
\end{align}
or
\begin{align} \label{eq:reverse}
\frac{g_\EHC F_6}{\Lambda_\EHC} \simeq 3 \ .
\end{align}
Even if we only make the very conservative assumption that $g_\EHC<1$, this result is not consistent with the expectation that $\Lambda_\EHC \gg F_6$.

In the above discussion we have ignored the running of the four-fermion coupling.
This coupling is presumed to be generated at the (high) EHC scale, where the estimate (\ref{eq:Gg}) is applicable.
The overlap factors are evaluated at the (low) hypercolor scale, and so the strength of the four-fermion coupling in \Eq{eq:ytsim} must be given at the hypercolor scale,
\begin{align} \label{eq:rung}
G(\Lambda_\HC) =
G(\Lambda_\EHC) \exp \left( -\int_{\Lambda_\HC}^{\Lambda_\EHC} \gamma(g_\HC(\mu)) \frac{d\mu}{\mu} \right) \ .
\end{align}
Here $g_\HC$ is the running gauge coupling of the hypercolor theory, while $\gamma$ is the anomalous dimension of the top-partner operator that
couples to the quark field via the four-fermion interaction.
(We neglect the effect of all the Standard Model gauge interactions, since they are presumed to be weak all the way from the EHC scale down to the hypercolor scale.)
If the anomalous dimension is large and negative over many energy decades, this running significantly enhances the four-fermion coupling  at the low scale.

Two considerations, however, prevent this enhancement.
First, our spectroscopy studies suggest that the model at hand is QCD-like, and not nearly conformal---the spectroscopy of slowly running theories, for example SU(3) with eight fundamental flavors, looks very different (see Refs.~\cite{DeGrand:2015zxa,Svetitsky:2017xqk} for reviews).
This implies that as we raise the energy scale above the hypercolor scale, the hypercolor coupling rapidly becomes perturbative.

Moreover, a one-loop calculation of the anomalous dimensions of the operators in \Eq{eq:4_fermion_operators} gives small values \cite{DeGrand:2015yna}.
This result has been corroborated by a higher-order perturbative calculation \cite{Pica:2016rmv}.
While the anomalous dimensions in the present theory have not been calculated non-perturbatively, these results indicate that the running of the four-fermion couplings in this model does not alleviate the problem exposed by \Eq{eq:reverse}.

We emphasize that even if this model were to exhibit large anomalous dimensions, our results for the overlap factor $Z$ indicate a self-consistency problem for the composite Higgs model.
The requirement that $y_t \simeq 1$ leads to \Eq{eq:100}, which can be rewritten in terms of the low-energy effective coupling as
\begin{align} \label{eq:Gconstraint}
G \simeq \frac{10}{F_6^2}\ ,
\end{align}
or if we rewrite $G \equiv 1 / \Lambda_G^2$ to put the coupling in terms of an energy scale, $\Lambda_G \simeq F_6 / 3$.  Even if a large enhancement is able to produce a $G$ of this magnitude from a weakly-coupled EHC theory, the energy scale associated with $G$ is well below the confinement scale, implying that this four-fermion coupling is strong at the hypercolor scale.
The basic assumption that we can describe the hypercolor sector in terms of a strongly-coupled gauge-fermion system which is weakly perturbed by the four-fermion couplings
is thus inconsistent; the dynamical effects of $G$ must be included from the outset.

\subsection{Summary and conclusions}

In this paper we have continued our lattice investigation of the \SU(4) gauge theory coupled simultaneously to fermions in the \4 and \6 representations.
This theory is closely related to a recent model of physics beyond the Standard Model, due to Ferretti, which contains a composite Higgs boson and a partially composite top quark.
In this scenario, the top quark couples linearly to heavy baryonic partners through four-fermion operators.
We calculated baryon overlap factors $Z_L$ and $Z_R$, defined in \Eq{eq:zb_definition}, which describe the overlap of the mixing operators with the top partner wave function.
We found that the overlap factors, while consistent with rough dimensional expectations, are about 20 times smaller than their analogues in QCD when measured in units of the decay constant of the Goldstone bosons.

Turning to phenomenological implications, we used our non-perturbative calculation of $Z_{L,R}$ to estimate the effective Yukawa coupling of the top quark.
Within our approximations, we find an inconsistency in the model.
Namely, the model is incompatible with the assumed separation of scales $\Lambda_\EHC \gg \Lambda_\HC$, if a realistic top Yukawa coupling is to be induced.
This result is independent of the precise value of the ratio of the hypercolor and electroweak scales.
In the case of two of our approximations it is difficult to estimate the precise systematic uncertainty.
These are the change in the number of sextet and fundamental flavors and the saturation of the low-energy constant in \Eq{LEC} by the lightest baryon.
Nevertheless, it is unlikely that improving on these approximations would reverse our negative conclusion.

This outcome is perhaps not surprising~\cite{Kaplan:1991dc,Panico:2015jxa}.
Conventional thinking on partial compositeness typically requires viable models to have near-conformal dynamics and mixing operators with large anomalous dimensions.
As discussed in Sec.~\ref{ssec:pheno} above, the model we study exhibits neither.  Still, the overlap factors $Z_{L,R}$ can only be determined reliably by
a non-perturbative lattice calculation.
Had these matrix elements been found to be much larger than their QCD counterparts, instead of much smaller as they actually turn out to be, the fate of the model might have been different.

Ferretti offered the $\SU(4)$ gauge theory with fermions in the \4 and \6 representations as a candidate model of new physics~\cite{Ferretti:2014qta}.
Our calculation of the effective Yukawa coupling of the top quark disfavors this particular model.
However, the $\SU(4)$ model was just one reasonably minimal choice within the broader classification of Ferretti and Karateev~\cite{Ferretti:2013kya,Ferretti:2016upr,Bizot:2016zyu,Belyaev:2016ftv,Agugliaro:2018vsu}.
Other models in their list remain interesting targets for lattice calculations.
Some of these models may exhibit near-conformal dynamics and may thus produce enhanced overlap factors.

Alternatively, starting from the current model
(or from the model of Ref.~\cite{Ferretti:2014qta}),
one can introduce additional massive fermions which are inert under
all the Standard Model symmetries, for the purpose of slowing down the running.
If the resulting theory has a ``walking'' coupling,
meaning that at $\Lambda_\HC$, where the coupling is strong,
the beta function is small, one might expect enhanced $Z$ factors
together with larger anomalous dimensions.  The latter will, in turn,
enhance the four-fermion couplings at the hypercolor scale;
the enhancement might stop at $G(\Lambda_\HC) \simeq g^2_{\rm eff}/\Lambda_\HC^2$
for some $g^2_{\textrm{eff}}\ll 1$, thereby avoiding the situation
of \Eq{eq:Gconstraint}.

\begin{acknowledgments}

We thank Maarten Golterman for many discussions.
We also thank Gabriele Ferretti and Kaustubh Agashe for correspondence.
Computations for this work were carried out with resources provided by the USQCD Collaboration,
which is funded
by the Office of Science of the U.S.\ Department of Energy; and with the Summit supercomputer, a
joint effort of the University of Colorado Boulder and Colorado State University, which is
supported by the National Science Foundation (awards ACI-1532235 and ACI-1532236), the
University of Colorado Boulder, and Colorado State University.
This work was supported in part by the U.S.\ Department of Energy under grant DE-SC0010005 (Colorado), and by the Israel Science Foundation under grants no.~449/13 and no.~491/17 (Tel Aviv).
Brookhaven National Laboratory is supported by the U.~S.~Department of Energy under contract DE-SC0012704.
This manuscript has been authored by Fermi Research Alliance, LLC under Contract No. DE-AC02-07CH11359 with the U. S. Department of Energy, Office of Science, Office of High Energy Physics.

\end{acknowledgments}

\appendix

\section{Matching to the low-energy theory} \label{app:matching}

The matching between the hypercolor and electroweak scales has been treated in detail for Ferretti's model in Ref.~\cite{Golterman:2015zwa}, while related calculations from a slightly more general perspective appear in Ref.~\cite{Golterman:2017vdj}.
In order to keep the present work self-contained, we provide an abbreviated discussion here.

Global symmetries together with matter content define an effective field theory.
As discussed in Sec.~\ref{sec:symmetries}, the structure of spontaneously broken global symmetries in the hypercolor sector of Ferretti's model is
\begin{align}
\left(\frac{\SU(5)}{\SO(5)} \right)
	\times \left( \frac{\SU(3) \times \SU(3)^\prime}{\SU(3)_{V}} \right)
	\times \left( \frac{\U1_B \times \U1_A}{\U1_B} \right),
\end{align}
and each broken symmetry in this product gives rise to a nonlinear field in the effective low-energy theory.
We denote the nonlinear field associated with the sextet fermions as $\Sigma = e^{2i \Pi/f}$.
In total, this nonlinear field describes 14 Goldstone bosons.
Four of them are identified as a composite Higgs doublet $H = (H_+, H_0)$.  Identifying the Standard Model's $\SU(2)_L \times \SU(2)_R$ with the $\SO(4)$ subgroup in the upper-left corner, the concrete embedding within the $\SU(5)/\SO(5)$ coset is $\Pi = H + H^\dagger + \cdots$ where \cite{Ferretti:2014qta}
\begin{align} \label{eq:hspurion}
H = \begin{pmatrix}
0 & 0 & 0 & 0 & -\frac{i}{\sqrt{2}}H_+ \\
0 & 0 & 0 & 0 & \frac{1}{\sqrt{2}}H_+ \\
0 & 0 & 0 & 0 & \frac{i}{\sqrt{2}}H_0 \\
0 & 0 & 0 & 0 & \frac{1}{\sqrt{2}}H_0 \\
-\frac{i}{\sqrt{2}}H_+ & \frac{1}{\sqrt{2}}H_+ & \frac{i}{\sqrt{2}}H_0 & \frac{1}{\sqrt{2}}H_0 & 0
\end{pmatrix}.
\end{align}
The broken symmetries in the fundamental sector and the conserved $\U1_A$ sector yield additional nonlinear fields, $\Omega$ and $\Phi$, which are discussed in Refs.~\cite{Golterman:2015zwa,Golterman:2017vdj}.
Because they play no role in the induced Yukawa couplings, we set them equal to unity for the rest of our discussion.

In order to derive the interactions between top sector and the composite Higgs field one begins by embedding the third-generation quark fields $q_L=(t_L,b_L)$ and $t_R$ into spurions transforming as the ${\bf 5}$ or ${\bf \bar{5}}$ of SU(5).
Since both the ${\bf 5}$ and the ${\bf \bar{5}}$ collapse to the defining
representation of SO(5), this determines the embedding to be
\begin{align} \label{eq:fspurions}
T_L = \frac{1}{\sqrt{2}}
		\begin{pmatrix}
		    ib_L\\
		    b_L\\
		    i t_L\\
		    -t_L\\
		    0
		\end{pmatrix}
		 ,\qquad
T_R = \begin{pmatrix}
			0\\
			0\\
			0\\
			0\\
			i t_R
		\end{pmatrix}.
\end{align}

The extended hypercolor dynamics induce effective four-fermion interactions at the hypercolor scale:
\begin{align} \label{eq:HC_eft}
V_\text{top}^\text{HC} =
	   G_R \bar{T}_L B_R
	+ G_L \bar{T}_R B_L
	+ \text{h.c.} .
\end{align}
The operators $B_{R,L}$ are top-partner baryon fields of definite chirality, which transform as the ${\bf 5}$ and ${\bf \bar{5}}$ of SU(5), respectively.
Integrating out all the heavy hypercolor states, including the top partners, produces an effective Lagrangian coupling the third-generation quark fields to the Goldstone bosons of the $\SU(5)/\SO(5)$ coset.
To leading order in the power counting of the low-energy theory, the effective Yukawa terms are
\begin{align}
V_\text{top}^\text{EW}
	= \mu_L \bar{T}_R \Sigma^* T_L
	+ \mu_R \bar{T}_L \Sigma T_R ,
\end{align}
where the coefficients $\mu_L$ and $\mu_R$ are low-energy constants.
The key physical point is that left- and right-handed quarks couple with an insertion of the nonlinear field $\Sigma$.

We now match the low-energy effective theory and the hypercolor theory to determine the relationship between the low-energy constants $\mu_{L,R}$ and
the four-fermion couplings $G_{L,R}$.
At the level of the partition functions, the matching amounts to equating the functional derivatives
\begin{align} \label{eq:eft_matching_condition}
\frac
	{\partial^2 \log \mathcal{Z}^\text{EW}}
	{\partial T_L \partial \bar{T}_R}
=
\frac
	{\partial^2 \log \mathcal{Z}^\text{HC}}
	{\partial T_L \partial \bar{T}_R} ,
\end{align}
(and similarly for $R \leftrightarrow L$).
The result is
\begin{align}
\mu_L P_L &= - G_L G_R P_L S_B(0) P_L , \\
\mu_R P_R &= - G_L G_R  P_R S_B(0) P_R ,
\end{align}
where
\begin{align}
S_B(p) \equiv \int d^4x\, e^{ipx} \left \langle B(x) \bar{B}(0) \right \rangle ,
\end{align}
and the four-component baryon field is $B = B_L + B_R$.
The appearance of the baryon two-point function corresponds to the fact that the functional derivatives on the right-hand side of \Eq{eq:eft_matching_condition} isolate two insertions of the interactions in \Eq{eq:HC_eft}.
For $p_\mu=0$, one expects that $S_B(0)$ is proportional to the identity in Dirac space.
Correspondingly, with slight abuse of notation, $\mu_L = \mu_R = -G_L G_R S_B(0)$.

The baryon two-point function $S_B(p)$ contains power-law divergences that originate when the baryon and anti-baryon are at the same point.
However, $P_L S_B(p) P_L$ and $P_R S_B(p) P_R$ are order parameters for $\SU(5)/\SO(5)$ symmetry breaking, and therefore their power divergences must be proportional to a positive power of the sextet fermion mass $m_6$.
In this way, all power-law divergences vanish in the sextet-chiral limit, $m_6 \rightarrow 0$.
On the lattice, avoiding such power divergences requires a fermion formulation with some chiral symmetry.
Because we have used Wilson fermions in our dynamical fermion simulations, this necessarily leads to a more complicated, mixed-action setup. We leave a direct study of $S_B(0)$ to a future project.

Instead, we conduct the following more modest calculation.
We expect the two-point function to be dominated by the lowest-lying baryon which couples to the operators $B_L$ and $B_R$.
Inserting a complete set of states reveals that
\begin{align}
\label{LEC}
S_B(0) = \frac{Z_L Z_R}{2 M_B} + \dots,
\end{align}
where $M_B$ is the mass of the lightest top-partner state and the dots denote contributions from excited states.
The factors $Z_L$ and $Z_R$ describe the overlap of the local chiral operators $B_L$ and $B_R$ with the baryon state.
They are the target of our lattice calculation and are defined in \Eq{eq:zb_definition}.

To see what mass is induced for the top quark, we set the Higgs to its vacuum expectation value $v$ and the spurion fields to their Standard Model values.
Using the concrete embeddings given in \Eq{eq:hspurion} and \Eq{eq:fspurions}, one readily discovers that
\begin{align}
 	\mu_L \bar{T}_R \Sigma^* T_L
 	+\mu_R \bar{T}_L \Sigma T_R
	&\longrightarrow
		-m_t \bar{t}t \ ,
\end{align}
where $t = t_L + t_R$ is the Standard Model's top quark, and the top quark mass is
\begin{align}
m_t =	\frac{G_L G_R S_B(0)}{\sqrt{2}} \sin \left( \frac{2\sqrt{2}v}{F_6} \right)
		\approx	G_L G_R
		\frac{Z_L Z_R}{M_B} \frac{v}{F_6} \ .
\end{align}
The right-most expression is valid after making the approximation of saturating the baryon two-point function by the lowest state, \emph{cf.} \Eq{LEC}, as well as the small-angle approximation $v/F_6 \ll 1$.
We can also infer the top quark's effective Yukawa coupling:
\begin{align}
y_t \approx G_L G_R \frac{Z_L Z_R}{M_B F_6} \ .
\end{align}

\section{Data tables \label{app:data_tables}}

This brief appendix collects tables of our simulation parameters and results for the baryon spectrum.
In order to keep the discussion self-contained, several of the tables have been reproduced from~\cite{Ayyar:2018zuk}.
Table~\ref{table:ensembles} lists the ensembles used in our calculation, while Table~\ref{table:fermion_masses} gives the fermion masses and Wilson flow scales for these ensembles.
Table~\ref{table:raw_overlaps} contains the results for the renormalized overlap factors.

\begin{table}[h]
\centering
\setlength{\tabcolsep}{10pt} 
\begin{tabular}{clllc}
\toprule
Ensemble	&  $\beta$ & $\kappa_4$ & $ \kappa_6$  & Configurations \\
\hline
1  &  7.25 &  0.13095 &  0.13418 &  61 \\
2  &  7.25 &  0.13147 &  0.13395 &  71 \\
3  &  7.30 &  0.13117 &  0.13363 &  61 \\
4  &  7.30 &  0.13162 &  0.13340 &  71 \\
5  &  7.55 &  0.13000 &  0.13250 &  84 \\
6  &  7.65 &  0.12900 &  0.13080 &  49 \\
7  &  7.65 &  0.13000 &  0.13100 &  84 \\
9  &  7.75 &  0.12800 &  0.13100 &  84 \\
10 &  7.75 &  0.12900 &  0.13080 & 54 \\
\toprule
\end{tabular}
\caption{
        The ensembles used in this study.
        All ensembles have volume $V = N_s^3 \times N_t = 16^3 \times 32$.
        The numbering of the ensembles matches that of Ref.~\cite{Ayyar:2018zuk}; we have dropped ensembles 8, 11, and~12 because fits to propagators involving point operators were unsuccessful.
	More details relating to these ensembles may be found in Refs.~\cite{Ayyar:2017qdf} and~\cite{Ayyar:2018zuk}.
}
\label{table:ensembles}
\end{table}

\begin{table}[h]
\centering
\setlength{\tabcolsep}{10pt} 
\begin{tabular}{clll}
\toprule
Ensemble         & $t_0/a^2$ & $\hat{m}_4$ & $\hat{m}_6$ \\
\hline
1  &  1.093(9) &  0.0422(7) &   0.020(1) \\
2  &  1.135(9) &   0.028(1) &   0.025(1) \\
3  &   1.13(1) &  0.0345(8) &   0.032(1) \\
4  &  1.111(9) &  0.0228(6) &  0.0381(8) \\
5  &   1.85(2) &   0.050(1) &   0.034(1) \\
6  &  1.068(5) &   0.082(1) &  0.0896(8) \\
7  &   1.46(2) &   0.046(2) &   0.080(2) \\
9  &   1.56(1) &   0.108(1) &   0.071(1) \\
10 &   1.75(2) &   0.073(2) &   0.077(2) \\
\toprule
\end{tabular}
\caption{
        Fermion masses and flow scales.
        Fermion masses are defined via the axial Ward identity.
        Measurement of these quantities is described in Ref.~\cite{Ayyar:2017qdf}.
}
\label{table:fermion_masses}
\end{table}

\begin{table}[h]
\centering
\setlength{\tabcolsep}{10pt} 
\begin{tabular}{cll}
\toprule
Ensemble 	& $\hat{Z}_R$	&	$\hat{Z}_L$	\\
\hline
	1 &   0.00064(6) &   0.00063(8) \\
        2 &   0.00063(9) &   0.00069(7) \\
        3 &   0.00070(6) &   0.00070(8) \\
        4 &   0.00075(5) &   0.00076(6) \\
        5 &   0.00108(7) &  0.00111(10) \\
        6 &   0.00111(8) &   0.00112(8) \\
        7 &   0.00114(7) &   0.00117(9) \\
        9 &  0.00143(13) &   0.00142(7) \\
       10 &   0.00133(9) &   0.00135(6) \\
\toprule
\end{tabular}
\caption{
	Results for the renormalized overlap factors $\hat{Z}_{L,R}^{\MSbar}(\mu=1/a)$ associated with the operators $B_{L,R}$.
	\label{table:raw_overlaps}
}
\end{table}

\section{Lattice spectroscopy \label{app:spectro}}

As with a general baryon interpolating field in QCD, the operator $\Lambda$ of \Eq{eq:lambda_operator} couples strongly to states of two different parities.
One of these states is the desired ground state, while the other is an excited state.
The mass and amplitude of the individual states are difficult to disentangle when both states are present.
The inclusion of a parity projection operator in \Eq{eq:c_ps} and \Eq{eq:c_ss} isolates each state.
On a lattice of infinite temporal extent, the correlator with $P_+$ would couple to the ground state only; the correlator with $P_-$ would couple to the excited state only.
The amplitudes $Z_{L,R}$ are defined as the overlap factors between the ground state and the point operators.

On a lattice of finite temporal extent, contributions are present from a backward-propagating state of opposite parity, even with the inclusion of the explicit parity projection $P_\pm$~\cite{Leinweber:2004it}.
A benefit of this fact is that \emph{both} projections contain information about the ground state.
Our analysis follows common practice in lattice baryon spectroscopy and combines the two projected correlation functions $C_{+}(t)$ and $C_{-}(N_t -t)$.
In this way, we obtain a single smeared-source, point-sink correlator which decays exponentially until $t \approx N_t/2$ and with amplitude $Z_B Z_\Lambda$.

In Sec.~\ref{ssec:correlation_functions}, we showed that pairs of correlation functions are related by discrete symmetry according to \Eq{eq:parity_relation_correlators}.
We have verified that our code which computes the correlations function satisfies \Eq{eq:parity_relation_correlators} to machine precision in free field theory.
We have also verified that the correlation functions in our simulations satisfy this relation with good agreement.
Our analysis also combines the correlations functions related by discrete symmetry.

Overall, on a configuration-by-configuration basis and before constructing any correlation matrix, we combine the correlation functions associated with $B_R$ and $B^\prime_L$:
$C_{+}^{B_R \bar{\Lambda}}(t)$,
$C_{+}^{B^\prime_L \bar{\Lambda}}(t)$,
$C_{-}^{B_R \bar{\Lambda}}(N_t - t)$, and
$C_{-}^{B^\prime_L \bar{\Lambda}}(N_t - t)$.
When combining the correlation functions, one must take care to mind overall signs [cf. \Eq{eq:parity_relation_correlators}].
Similarly, we combine the correlation functions with the sinks operators $B_L$ and $B^\prime_R$.

We then fit the correlation functions in Eqs.~(\ref{eq:c_ps}) and~(\ref{eq:c_ss}) to a single decaying exponential instead of a hyperbolic function, neglecting the region with $t \gtrsim N_t/2$ which remains contaminated by the excited state of opposite parity.
For each correlator, we use the fitting procedure described in Appendix~B of Ref.~\cite{Ayyar:2017qdf}.
In particular, we include a systematic uncertainty stemming from the choice of the initial and final times $[\tmin, \tmax]$.

We use the publicly available Python packages \texttt{lsqfit}~\cite{Lepage:lsqfit} and \texttt{gvar}~\cite{Lepage:gvar} for nonlinear fitting and classical error propagation.
When computing ratios of quantities derived from different fits, we use single-elimination jackknife to propagate errors including correlations.

\begin{figure}[t]
\includegraphics[width=1.0\textwidth]{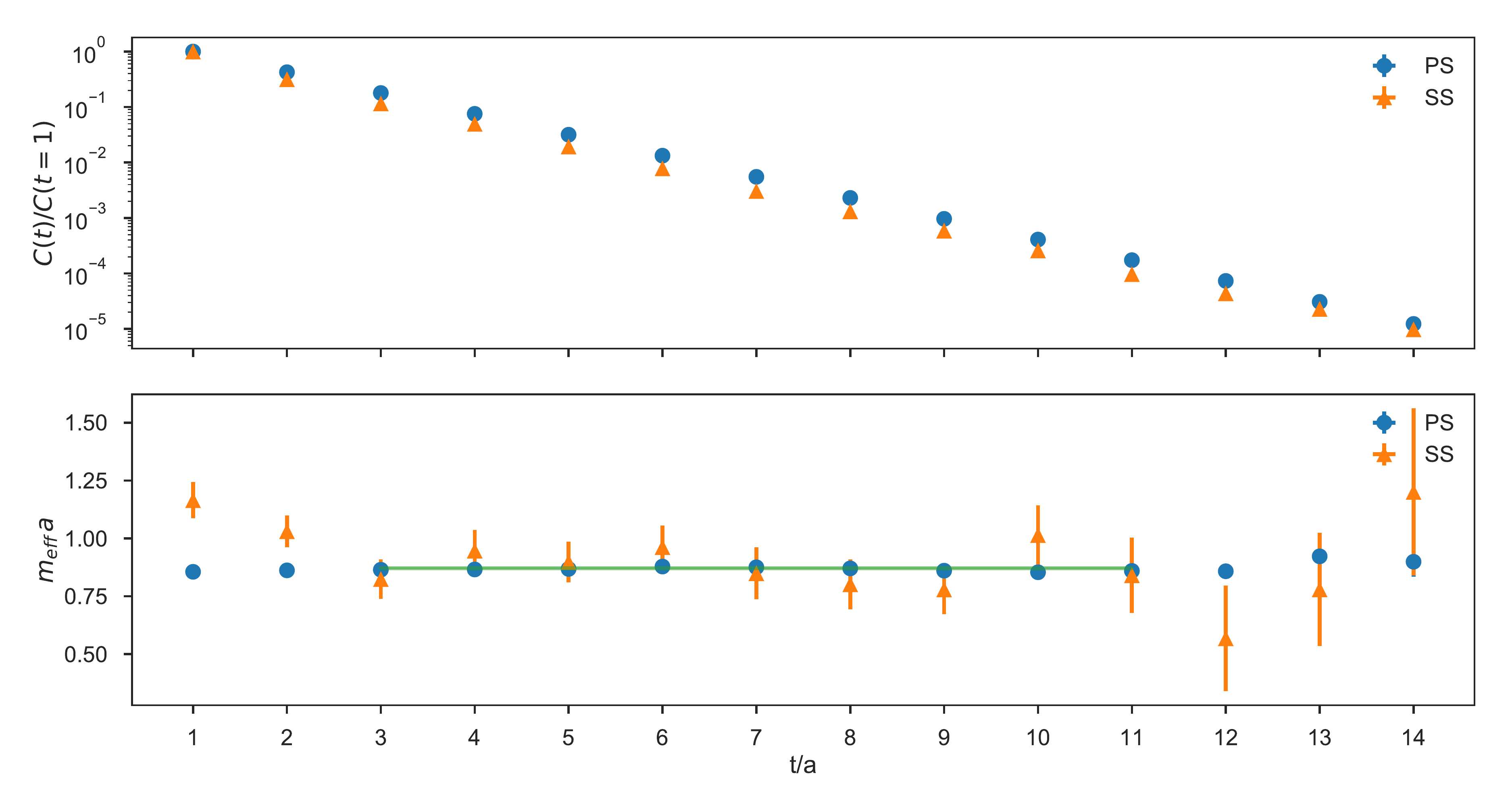}
\caption{
	The correlation functions and effective masses of \Eq{eq:c_ps} and \Eq{eq:c_ss} used to extract $Z^{RR}$ for Ensemble 10.
	Both correlation functions exhibit strong signals throughout the fit region.
	The green band indicates mass of the best-fit result; the width indicates statistical uncertainty only.
	\label{fig:correlator}
}
\end{figure}

\begin{figure}[t]
\includegraphics[width=1.0\textwidth]{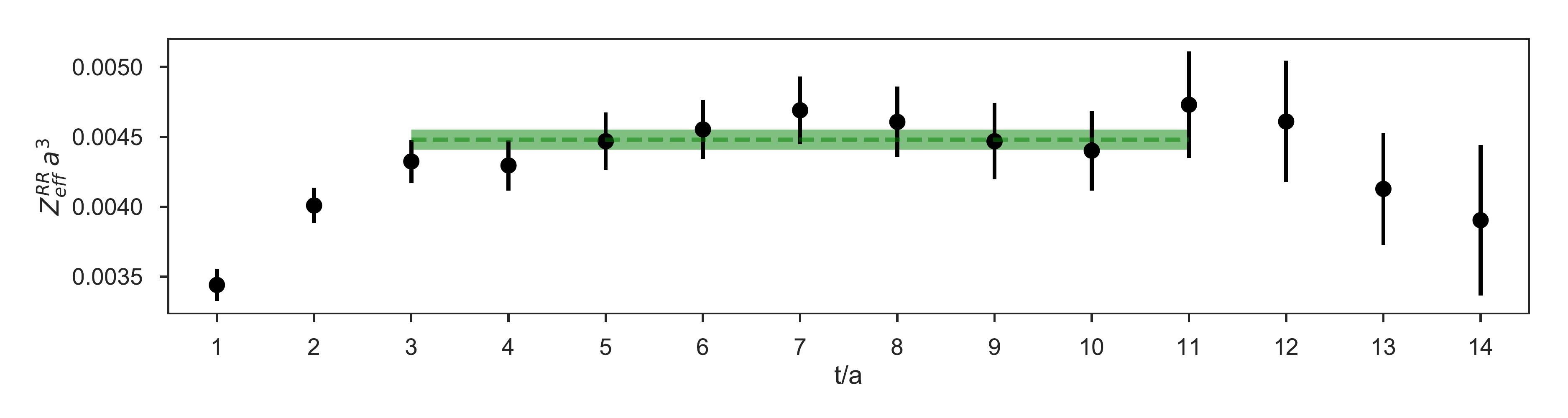}
\caption{
	The effective amplitude $Z_\text{eff}^{RR} a^3$ of \Eq{eq:zeff} for Ensemble 10.
	The green band indicates the best-fit result and fit range; the width indicates statistical uncertainty only.
	\label{fig:effective_amp}
}
\end{figure}

\begin{figure}[t]
\includegraphics[width=1.0\textwidth]{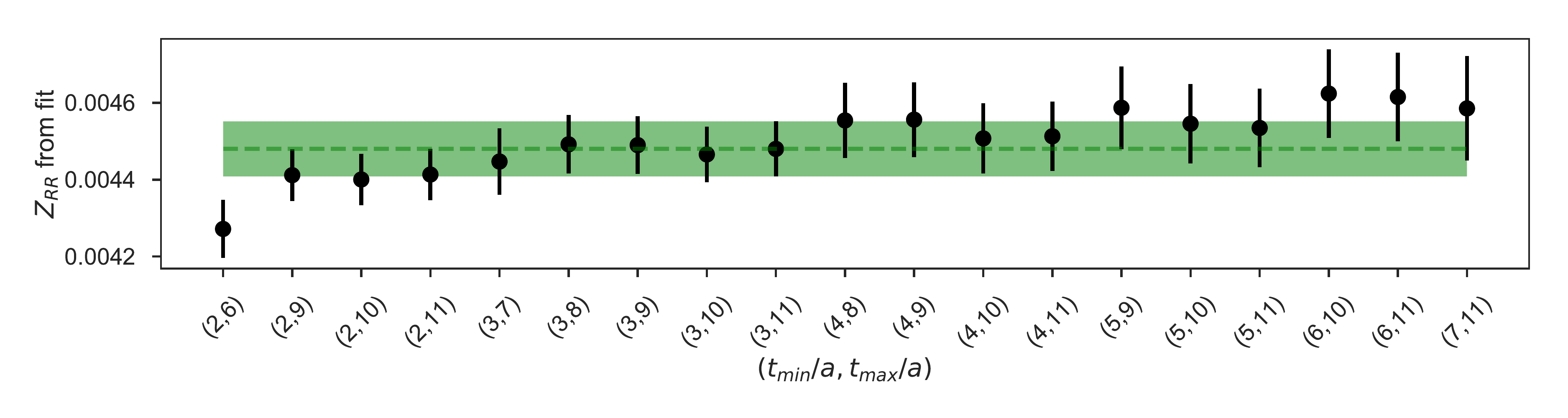}
\caption{
	Values for the amplitude $Z^{RR} a^3$ emerging from other candidate fits for Ensemble 10.
	The green band indicates the best-fit result; the width indicates statistical uncertainty only.
	\label{fig:grid_search}
}
\end{figure}

To demonstrate the stability of our fits, we now show some illustrative results from Ensemble 10; the other ensembles are similar.
Figure~\ref{fig:correlator} shows the correlation functions and effective masses of \Eq{eq:c_ps} and \Eq{eq:c_ss} used to determine $Z^{RR}$.
The effective mass is computed according to
\begin{align}
m_\text{eff}(t)a = \arcosh \left( \frac{C(t+1) + C(t-1)}{2 C(t)} \right).
\end{align}
Both correlation functions exhibit strong signals throughout the fit region.
The green band is the result of our best fit and extends across the fit region chosen according to the procedure in Appendix~B of Ref.~\cite{Ayyar:2017qdf}.
In order to achieve strong signals and flat effective masses we tuned the gaussian smearing radius on each ensemble, as described in Ref.~\cite{Ayyar:2018zuk}. 
As a non-trivial check of our results, we verified that the masses in this study agreed statistically with our previous work using different interpolating fields~\cite{Ayyar:2018zuk}.
Figures~\ref{fig:effective_amp} and~\ref{fig:grid_search} demonstrate the stability of our determination of $Z^{RR}$.
Figure~\ref{fig:effective_amp} shows the effective amplitude $Z_\text{eff}^{RR}$, with the black points constructed according to
\begin{align} \label{eq:zeff}
Z_\text{eff}(t) 	&= \frac{C^{PS}(t)}{\sqrt{C^{SS}(t)} e^{-m_B t/2}},
\end{align}
with $m_B$ taken from the best fit.
Similar effective amplitudes arise frequently in lattice QCD studies of flavor physics (for one such example, see \cite{Bailey:2008wp}).
The signal for the amplitude is stable and consistent with our best-fit result.
The outliers at early times likely contain excited-state contamination.
At large times not included in the best fit, the effective amplitude remains statistically consistent with the best-fit result.
Figure~\ref{fig:grid_search} shows results for the amplitude $Z^{RR}$ coming from other candidate fits using different fit windows.
The figure only shows successful fits with $\chi^2/\text{DOF} \lesssim 2.0$.
The figure demonstrates that the fit result for the amplitude is robust to the choice of fitting window.
Following the procedure of Ref.~\cite{Ayyar:2017qdf}, our final results also include a conservative systematic error to account for any possible bias arising from the choice of the fit window.
The systematic error for $Z^{RR}$ is 0.00015 for this ensemble.
The general features of the analyses are similar for $Z^{LR}$ and for the other ensembles.

\section{Normalization and Renormalization \label{app:renormalization}}

The mass of the top partner is a renormalization group invariant quantity.
The amplitude $Z_B$, however, depends on the scale and must therefore be renormalized in order to make contact with continuum physics.
This process consists of a couple of steps, which we now describe.

First, Wilson fermions carry a different overall normalization from continuum fermions.
Correcting this discrepancy amounts to multiplying each fermion field by a factor of $\sqrt{1- 3\kappa_r/4\kappa_{rc}}$, where $\kappa_r$ is the hopping parameter and $\kappa_{rc}$ is its critical value~\cite{Lepage:1992xa}.
The subscript $r$ denotes the representation of the fermion.
Therefore, a baryon operator of the form $\mcal{O}_B \sim Qqq$ acquires the following normalization
\begin{align}
\mcal{O}_B
	\rightarrow \sqrt{1- \frac{3}{4} \frac{\kappa_\6}{\kappa_{\6c}}} \times \left(1- \frac{3}{4} \frac{\kappa_\4}{\kappa_{\4c}} \right) \mcal{O}_B
	\equiv N(\kappa_\4,\kappa_\6) \mcal{O}_B.
\end{align}

Second, we require a matching coefficient $Z(\text{latt}\rightarrow\MSbar)$ which converts a lattice regulated matrix
element into its dimensionally-regulated analog in the \MSbar~ scheme.
At one loop, one finds
\begin{align}
Z(\text{latt}\rightarrow\MSbar) = 1 + \frac{\alpha_{\MSbar}(q_B^*)}{4\pi} \mcal{Z}, \label{eq:lattice_to_ctm}
\end{align}
where $\mcal{Z}$ is the difference between the (finite portion of the) \MSbar~ integral in $4-2\epsilon$ dimensions and a corresponding integral in lattice perturbation theory
\begin{align}
\mcal{Z} = I_{\MSbar}^\text{finite} - I_\text{lattice}.
\end{align}
$I_{\MSbar}^\text{finite}$ and $I_\text{lattice}$ are the results of one-loop calculations in continuum and lattice perturbation theory, respectively.
Ref.~\cite{DeGrand:2015yna} carried out the relevant calculation in continuum perturbation theory.
A standard but rather technical computation along the lines of Ref.~\cite{DeGrand:2002va} delivers the result in lattice perturbation theory.
Appendix~D of Ref.~\cite{Ayyar:2017qdf} contains more details relevant to the calculation in the present \SU(4) system.

This calculation makes a simplifying approximation. It can be illustrated by looking at a vertex correction.
Think of a vertex operator as $\bar \psi_\alpha \Gamma_i \psi_\beta$ for Dirac matrix $\Gamma_i$ and color factors $\alpha$ and $\beta$ on the spinors and write this quantity as $\Gamma_i$.
The one loop correction to $\Gamma$ is
\begin{align}
V^\Gamma =  K_0 \Gamma  +
 K_1 \gamma_\mu \Gamma\gamma_\mu
+K_2 \gamma_\mu \gamma_\nu \Gamma \gamma_\nu \gamma_\mu
+ \dots.
\label{eq:vadesire}
\end{align}
where the $K_i$'s are individual lattice integrals which can be computed by projecting integrands onto elements of the Clifford algebra.

The Wilson and clover actions have only $K_0$, $K_1$, and $K_2$ nonzero.
 The overlap action only has nonzero $K_0$ and $K_2$ terms.
The continuum calculation with massless fermions only has nonzero $K_2$.
More complicated actions could have additional terms.
The $K_1$  term is responsible for  ``bad'' operator mixing into opposite-chirality operators.
It is the source of the biggest artifacts in lattice calculations of four fermion operators like $B_K$ with Wilson-type quarks.
However, lattice studies using clover-improved Wilson quarks have successfully suppressed this mixing using smearing~\cite{DeGrand:2002va,Durr:2011ap}.
In the present study we use clover-improved Wilson fermions with nHYP smearing, which may be therefore expected to reduce mixing.

Calculation shows that with nHYP clover fermions and the usual Wilson gauge action, $K_0=4.38$, $K_1= -0.02$, $K_2=-0.47$.
The tiny value of $K_1$ suggests that we should not worry about lattice induced mixing effects and just take $K_1=0$.
For comparison, the value of $K_1$ with thin links is nearly 20 times larger~\cite{DeGrand:2002va}.
This allows us to quickly extend the mnemonic of \Eq{eq:vadesire} to all the one loop perturbative diagrams, even ones which are not so easily Fiertz-rotated into a product of a dressed current times an undressed.
The matching factor is the same for all four operators in \Eq{eq:4_fermion_operators}.
We find $\mcal{Z} = -4.83$.

\Eq{eq:lattice_to_ctm} contains a well-understood ambiguity: what are the correct value and scale for the running coupling?
Many reasonable solutions exist to this problem.
We elect to use the scheme due to Lepage and Mackenzie \cite{Lepage:1992xa}, which defines the coupling $\alpha_V$ from a non-perturbative measurement of the trace of the plaquette operator on each ensemble.
After converting this coupling to an $\MSbar$ value~\cite{Brodsky:1982gc,Hornbostel:2002af}, we run it to a momentum scale $q_B^\star a$ characteristic of the operator $\mcal{O}_B$.
Hornbostel, Lepage, and Morningstar provide a prescription for computing $q_B^\star a$ in lattice perturbation
theory~\cite{Hornbostel:2000ey,Hornbostel:2002af}.
Their procedure requires slight modification for operators with an anomalous dimension;
our precise technique is that of Ref.~\cite{DeGrand:2002va}.
We find $q_B^\star a =1.15$.

We remark that the values for $\mcal{Z}$ and for $q_B^\star$ agree (within the quoted digits) for the NDS action used in this study with the corresponding results using the Wilson gauge action.

Assembling all of our pieces,
\begin{align}
	Z^{\MSbar}_{B}(\mu = 1/a) =
		Z(\text{latt}\rightarrow\MSbar)
		N(\kappa_\4,\kappa_\6)
		Z^\text{lattice}_{B}(\mu = 1/a).
\end{align}
The quantity $Z^\text{lattice}_{B}$ is what emerges from the fits to lattice data; the physical quantity is $Z^{\MSbar}_{B}(\mu = 1/a)$.
In the ensembles of this study, the overall multiplicative factor is 
\begin{align}
Z(\text{latt}\rightarrow\MSbar) N(\kappa_\4,\kappa_\6) \simeq 0.12.
\end{align}

\bibliography{baryon_ff}
\end{document}